\documentclass[aps,prd,preprint,floatfix,superscriptaddress]{revtex4}

\usepackage{graphicx}

\def\ra{\rightarrow}

\def\bwt{\begin{widetext}}
\def\ewt{\end{widetext}}
\def\be{\begin{equation}}
\def\ee{\end{equation}}
\def\bea{\begin{eqnarray}}
\def\eea{\end{eqnarray}}
\def\bean{\begin{eqnarray*}}
\def\eean{\end{eqnarray*}}
\def\bary{\begin{array}}
\def\eary{\end{array}}
\def\bit{\begin{itemize}}
\def\eit{\end{itemize}}

\def\ra{\rightarrow}

\begin{document}

\setcounter{page}{0}
\thispagestyle{empty}

\preprint{ANL-HEP-PR-03-028, hep-ph/0304267}
\bigskip

\title{Differential Cross Sections for Higgs Boson Production 
at Tevatron Collider Energies}

\author{Edmond~L.~Berger}
\email[e-mail: ]{berger@anl.gov}
\affiliation{High Energy Physics Division, 
Argonne National Laboratory, Argonne, IL 60439}
\author{Jianwei~Qiu}
\email[e-mail: ]{jwq@iastate.edu}
\affiliation{Department of Physics and Astronomy, 
Iowa State University, Ames, IA 50011}

\date{\today}

\begin{abstract}
The transverse momentum $Q_T$ distribution is computed for inclusive 
Higgs boson production at $\sqrt{S} = 1.96$~TeV.  
We include all-orders resummation of large logarithms associated with 
emission of soft gluons at small $Q_T$.  We provide results for Higgs 
boson and $Z^*$ masses from $M_Z$ to 200~GeV.  
The relatively hard transverse momentum distribution for Higgs boson 
production suggests possibilities for improvement of the signal to 
background ratio.  

\end{abstract}


\maketitle


\section{INTRODUCTION}

Elucidation of the dynamics responsible for the breaking of 
electroweak symmetry is one of the primary goals in 
particle physics during this decade.  In the current run-II of the 
Fermilab Tevatron, the upgraded Collider Detector Facility (CDF) and 
D0 detectors will search for the neutral Higgs boson ($h$), the vehicle 
of symmetry breaking in the standard model (SM), as well as the Higgs 
bosons of the minimal supersymmetric standard model (MSSM).  The task 
is challenging.  The production cross sections and branching fractions 
into channels favorable for detection are relatively small, and the  
backgrounds from competing processes are large~\cite{Carena:2000yx}. 
 
The fully inclusive gluon-gluon fusion 
subprocess~\cite{Wilczek:1977zn,Graudenz:1992pv} $g g \ra h X$ 
supplies the largest cross section at Tevatron energies.  Its rate 
is a factor four or so greater than that of the next most important 
subprocess, associated production $q \bar{q} \ra h V X$, where 
$V = W, Z$.  
In the region of modest Higgs boson masses, $m_h < 135$~GeV, the 
combination of gluon fusion production with the dominant decay 
process $h \ra b \bar{b}$ should be attractive, but its promise 
is compromised by the overwhelming production of $b \bar{b}$ pairs 
from strong interactions background processes. For $m_h > 135$~GeV, the 
channel $h \ra W W^*$ becomes dominant, and gluon fusion followed by 
$h \ra W W^*$ provides a potential discovery mode at the Tevatron.  

Calculations of the expected differential cross sections for production 
of the signal and backgrounds are important for evaluation of the 
required measurement accuracies and detector performance.  Estimations 
of the expected transverse transverse momentum distributions can suggest 
selections in this variable that should improve background rejection.  
In this paper, we concentrate on $g g \ra h X$ and discuss the behavior 
of the Higgs boson transverse momentum distribution in the region of small 
and intermediate values of $Q_T$.  We expect our 
predictions to be directly pertinent in the region $m_h > 135$~GeV, but 
we also entertain optimism that experimenters may find clever ways
to make use of the dominant channel $g g \ra h X$ in the region 
$m_h < 135$~GeV.  

In the gluon fusion process, production occurs through triangle loops of 
colored (s)particles that couple to the Higgs boson and to gluons.  In the 
SM, the most relevant contribution is from a loop of top ($t$) 
quarks. The coupling of gluons is simplified in the limit of large top 
quark mass $m_t$~\cite{Ellis:1975ap}.  The $m_t \rightarrow \infty$ 
approximation is valid to an accuracy of 
$\sim 5$\% for $m_h \le 2 m_t$~\cite{Dawson:1990zj}.  Within this approach, 
the total cross section for $g g \rightarrow h X$, 
is known to next-to-next-to-leading order (NNLO)
accuracy~\cite{Harlander:2002wh,Ravindran:2003um}. 
The large $m_t$ approximation serves well also for the transverse momentum 
distribution when $m_h < 2 m_t$ and the Higgs boson transverse momentum 
$Q_T$ is less than $m_t$~\cite{Ellis:1987xu}.  The next-to-leading order 
contributions are computed in 
Refs.~\cite{deFlorian:1999zd,Ravindran:2002dc,Glosser:2002gm}.  

When $Q_T$ is comparable to $m_h$, there is only one hard momentum scale in 
the perturbative expansion of the cross section as a function of the strong 
coupling $\alpha_s$, and fixed-order computations in perturbative QCD are 
expected to be applicable. However, in the region $Q_T \ll m_h$, where the 
cross section is greatest, the coefficients of the expansion in $\alpha_s$ 
depend functionally on logarithms of the ratio of the two quantities, 
$m_h$ and $Q_T$, $(\alpha_s/\pi) \ln^2(m_h^2/Q_T^2)$.
The relevant expansion parameter in the perturbative series 
is close to 1, and straightforward fixed-order perturbation 
theory is inapplicable for $Q_T \ll m_h$.  All-orders resummation is 
the established method for mastering the large logarithmic coefficients 
of the expansion in $\alpha_s$ and for obtaining well-behaved cross 
sections at intermediate and small 
$Q_T$~\cite{Collins:1981uk,Collins:1984kg,Ellis:1997ii}.  
Using renormalization group techniques, Collins, Soper, and 
Sterman (CSS)~\cite{Collins:1984kg} devised a $b$ space resummation formalism 
that resums all logarithmic terms as singular as $(1/Q_T^2)\ln^n(m_h^2/Q_T^2)$ 
when $Q_T \rightarrow 0$.  This formalism has been used widely for computations 
of the transverse momentum distributions of Higgs 
bosons~\cite{Catani:vd,Hinchliffe:1988ap,Kauffman:1991jt,Yuan:1991we,deFlorian:2000pr,Berger:2002ut,Bozzi:2003jy} 
and other processes.  

In this paper, we use the $b$ space resummation approach discussed in some 
detail in our Ref.~\cite{Berger:2002ut} to derive predictions for the $Q_T$ 
spectrum of Higgs boson production at the Fermilab Tevatron energy 
$\sqrt{S} = 1.96$~TeV.  We resum 
the large logarithmic terms associated with soft 
gluon emission to all orders in $\alpha_s$ obtaining well defined predictions 
for the full range of $Q_T$.  We employ expressions for the 
parton-level hard-scattering functions valid through first-order in $\alpha_s$, 
including contributions from the glue-glue, quark-glue, and quark-antiquark 
incident partonic subprocesses.  We present differential cross sections for 
masses of the Higgs boson that span the range of present interest in 
the SM, from $m_h = M_Z = 91.187$~GeV to $m_h = 200$~GeV. To illustrate 
interesting differences, we also provide results for the $Z$ boson.  
In this paper, we also show the integrated $Q_T$ distributions 
$\sigma(Q_T > Q_{Tmin})$ for rapidity within the interval $|y| < 1.0$.  This 
distribution indicates immediately what fraction of the Higgs boson cross section 
may be sacrificed if a selection is made on $Q_T$ for background rejection.  

Our predictions are presented and discussed in Sec.~II.  Conclusions and a 
discussion of background rejection are found in Sec.~III.  
%
\section{Predictions}
Our predictions are based on the formalism described in detail in our 
Ref.~\cite{Berger:2002ut}.  For the perturbative $A_g$ and $B_g$ functions 
in the Sudakov factor, we make an expansion valid through second order 
($n = 2$) in the strong coupling $\alpha_s$. For the short-distance 
coefficient functions $C_{a \rightarrow b}$ used to compute the modified 
parton densities, we retain the expansion through $n = 1$, {\em i.e.}, 
to first order in $\alpha_s$.  We remark that the functions 
$A_g$, $B_g$, and $C_{a \rightarrow b}$ have well behaved perturbative 
expansions, free of large logarithmic terms.  We use a next-to-leading 
order form for $\alpha_s(\mu)$ and next-to-leading order normal parton  
densities $\phi(x,\mu)$~\cite{Lai:1999wy}.  In the fixed-order 
perturbative expressions that enter the ``$Y$'' function, we select a 
fixed renormalization/factorization scale 
$\mu = \kappa \sqrt{m_h^2 + Q_T^2}$, with $\kappa = 0.5$.  For the 
resummed term, we take as our central value $\mu = c/b$, with 
$c=2e^{-\gamma_E} \approx 1$.  We examine sensitivity to the choice of $\mu$ by 
computing cross sections with other choices of this scale.  
The extrapolation into the non-perturbative 
region of large $b$ is accomplished with the form devised by Qiu and 
Zhang~\cite{Qiu:2000hf}.

\begin{figure}[ht]
\centerline{\includegraphics[width=9.0cm]{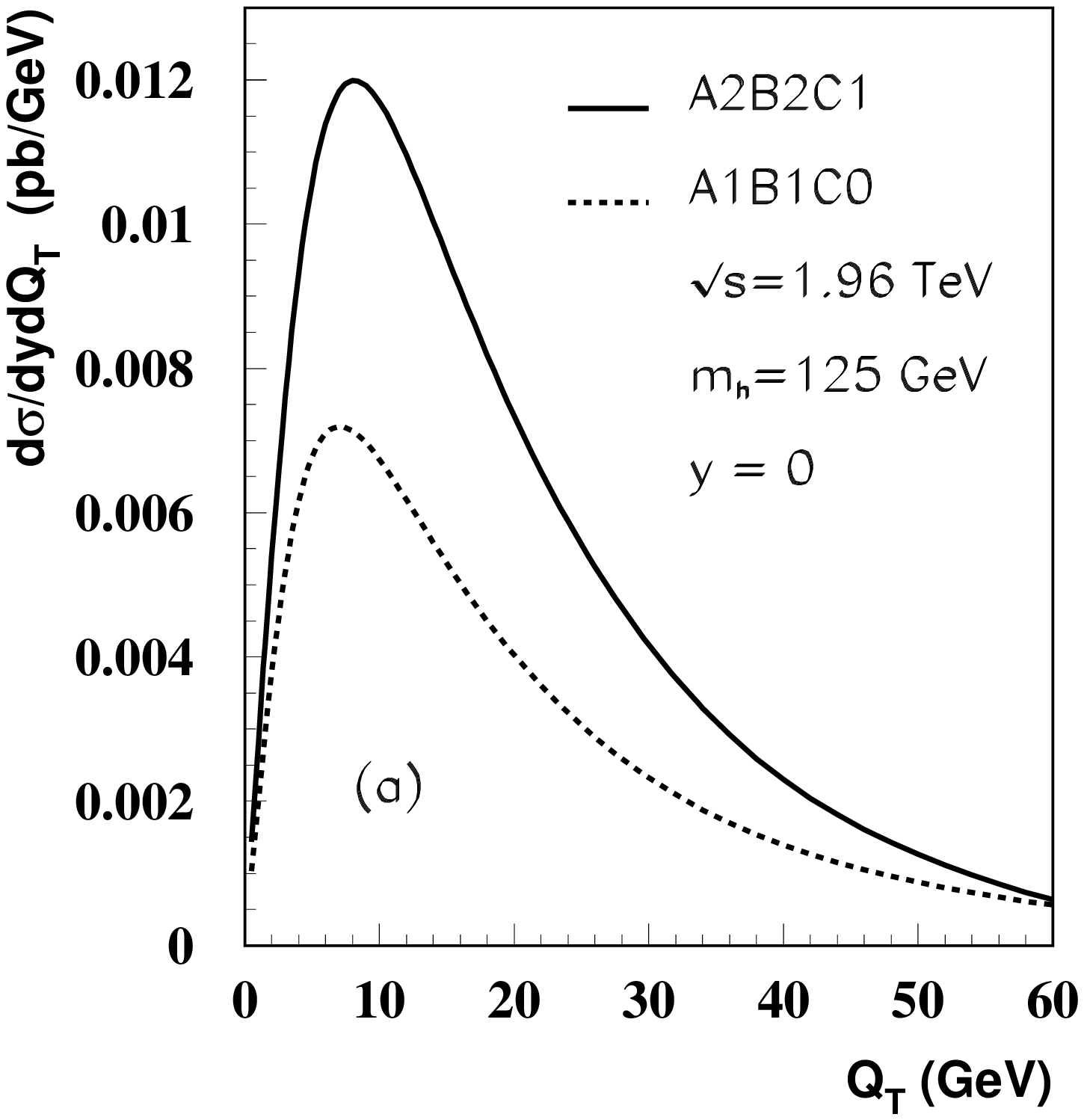}
\includegraphics[width=9.0cm]{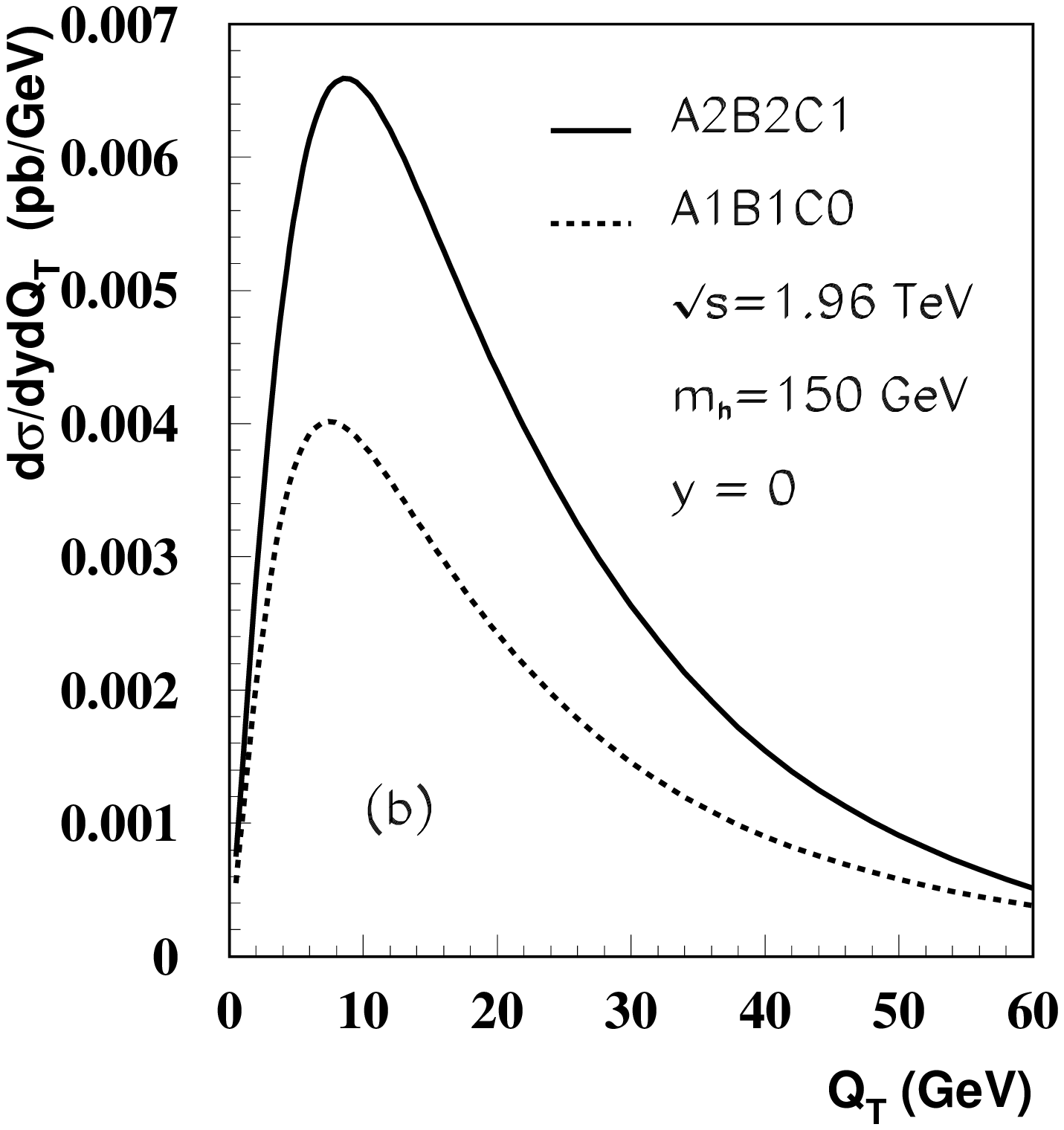}} 
\centerline{\includegraphics[width=9.0cm]{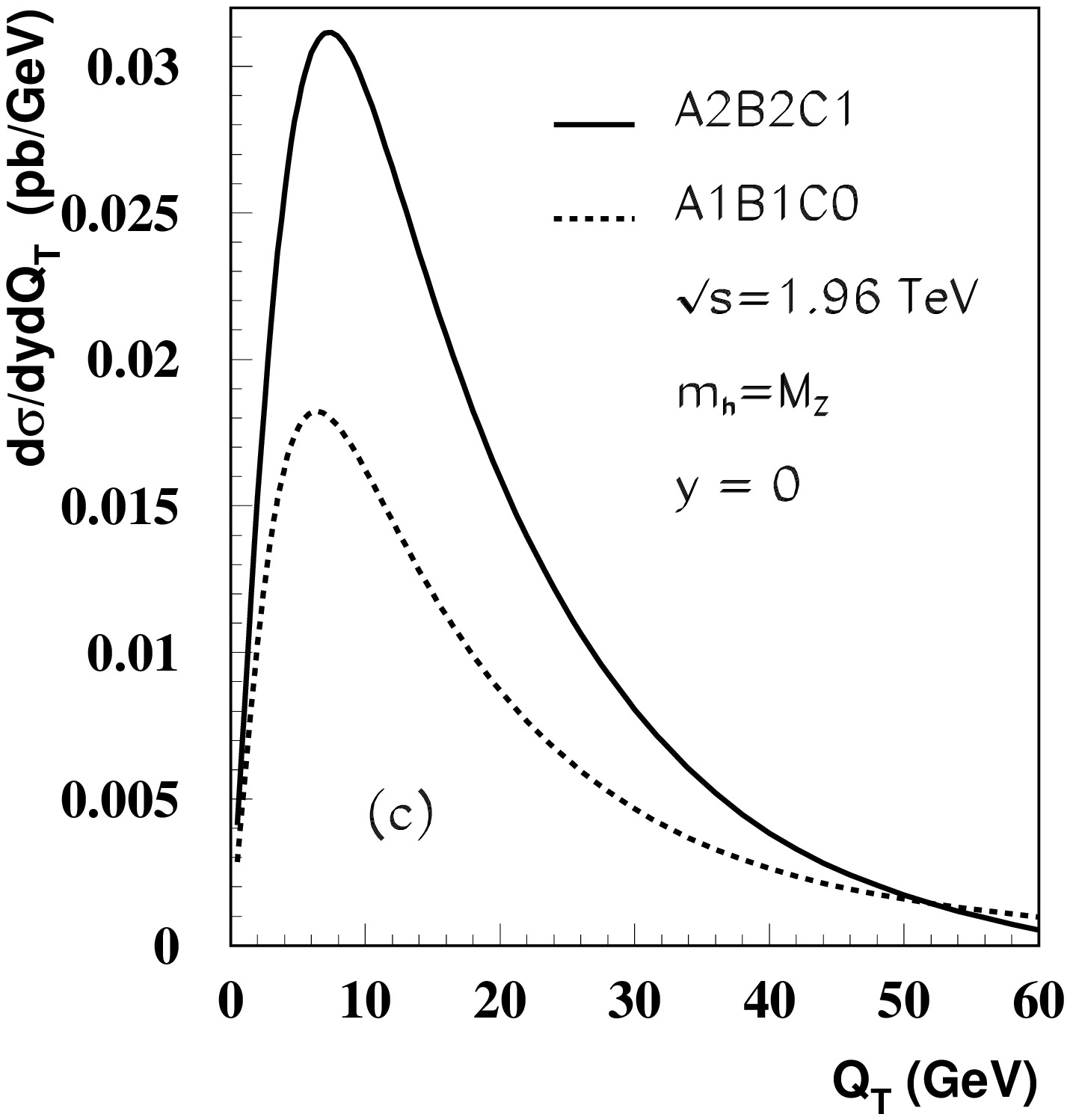}
\includegraphics[width=9.0cm]{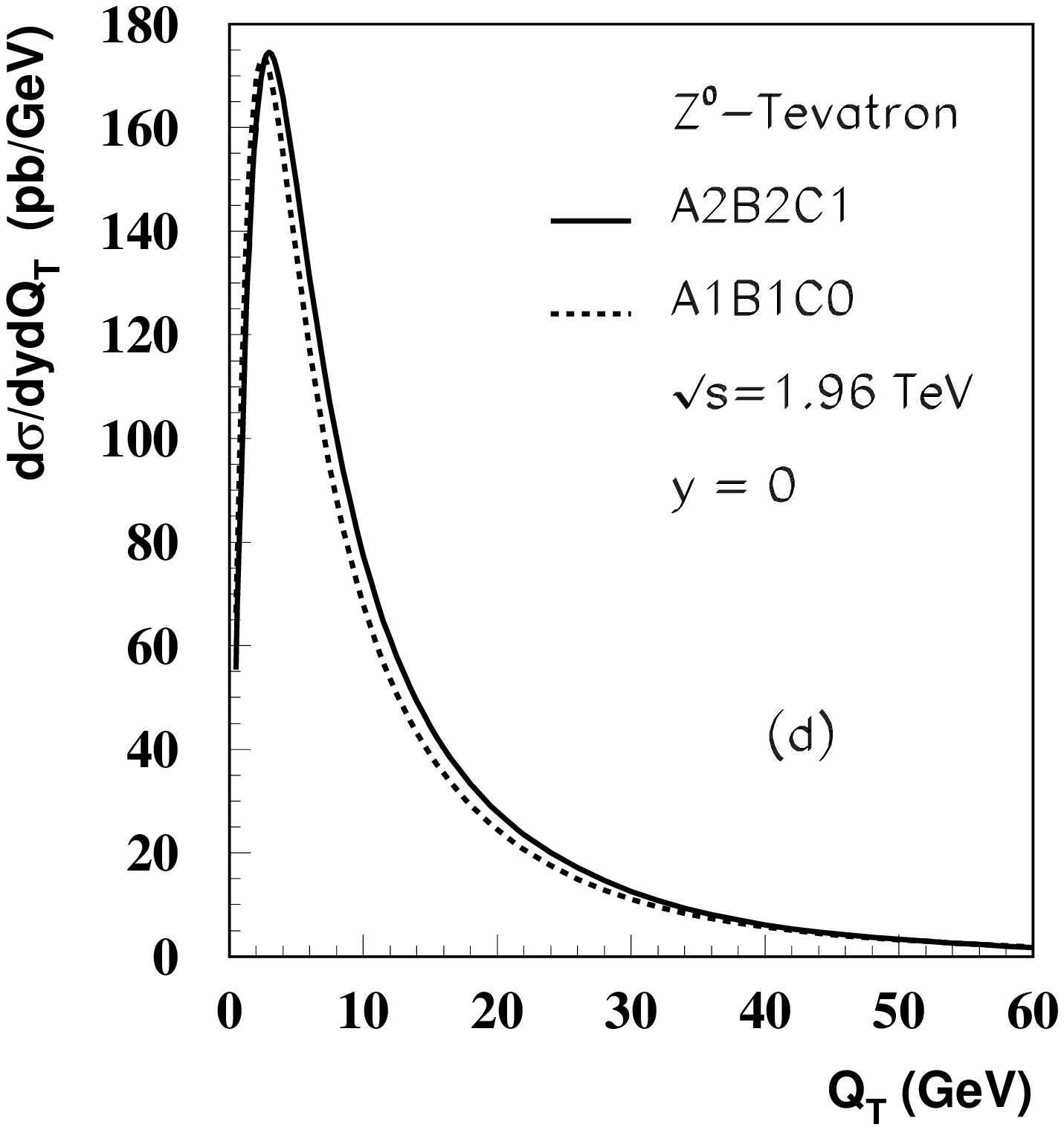}} 
\caption[]{\it Differential cross sections for Higgs boson production at 
  $\sqrt{S} = 1.96$~TeV at 
  (a) $Q=m_h=125$ GeV, (b) $Q=m_h= 150$ GeV, and (c) $Q= M_Z$.  
  The NLO prediction is shown as a solid line and the LO 
  prediction as a dashed line. In (d) we show analogous results 
  for $Z$ boson production.}
\label{fig:QTHiggs}
\end{figure}

In Fig.~\ref{fig:QTHiggs}, we show the predicted differential $Q_T$ 
distributions.  We present results for three choices of mass of the Higgs 
boson, $m_h = M_Z = 91.187$~GeV, $m_h = 125$ GeV, and $m_h = 150$~GeV 
(where the $W W^*$ decay channel is dominant).
In all cases, the solid lines represent the predictions of 
resummation at next-to-leading order (NLO) accuracy  
and, for comparison, the dashed lines show the results of 
resummation at leading-order (LO) accuracy~\cite{NLO-LO}.     
  
Two points are evident in the comparison of $Z$ boson and Higgs boson 
production, with $m_h = M_Z$, Fig.~\ref{fig:QTHiggs}(c) and (d).  The peak 
in the $Q_T$ distribution occurs at a 
smaller value of $Q_T$ for $Z$ production.  At $y = 0$, 
the curve peaks at $Q_T \sim 3.0$~GeV for $Z$ production and at 
$Q_T \sim 7.3$~GeV for Higgs boson production.  Second, the distribution 
is narrower for $Z$ production, falling to half its maximum 
by $Q_T \sim 9.0$~GeV, whereas the half-maximum for Higgs production is 
not reached until $Q_T \sim 21$~GeV. The physics behind these important 
differences is that the 
larger QCD color factors produce more gluonic showering in the glue-glue 
scattering subprocess that dominates inclusive Higgs boson production than 
in the fermionic subprocesses relevant for $Z$ production.  After all-orders 
resummation, the enhanced showering suppresses the large-$b$ (small $Q_T$) 
region more effectively for Higgs boson production.  

For Higgs boson production, there is a substantial difference in the 
predictions at NLO and at LO, but the differences are slight for $Z$ 
production.  The role of the $C$ function is responsible for these 
effects.  The first two coefficients in the expansion of $C$ are 
published~\cite{Kauffman:1991jt,Yuan:1991we,deFlorian:2000pr}. 
In powers of ($\alpha_s/\pi$), the $C^{(1)}$ 
coefficients for the $gg$ and $q \bar{q}$ cases are 
\begin{eqnarray}
C^{(1)}_{g\rightarrow g}(z) &=& \delta(1-z) \frac{1}{2}
  \left[ C_A \left(\frac{2\pi^2 + 11}{6} \right) 
        +\frac{\pi^2}{6}C_A
  \right] .
\nonumber \\
C^{(1)}_{q\rightarrow q}(z) &=& \delta(1-z) \frac{1}{2}
  \left[ C_F \left(\frac{2\pi^2 - 24}{6} \right) 
        +\frac{\pi^2}{6}C_F \right] + \frac{1}{2} C_F (1 - z) .
\label{css-coef}
\end{eqnarray}
In the $q \bar{q}$ case relevant for $Z$ production, there is a strong 
cancellation, $2\pi^2 - 24$, reducing the effect of the NLO 
contribution.  For Higgs boson production, the analogous 
term is additive, $2\pi^2 + 11$.  The constant terms $-24$ and $11$, 
respectively, are associated with virtual corrections at NLO and can 
be traced to the interference between the tree and one-loop diagrams. 
The  $2\pi^2 - 24$ 
term is effectively the finite piece of the one-loop correction to 
the $q \bar{q} \rightarrow Z$ vertex while the $2\pi^2 + 11$ term 
results from the finite piece of the one-loop correction to the 
effective $g g \rightarrow h$ vertex.  

\begin{figure}[ht]
\centerline{\includegraphics[width=9.0cm]{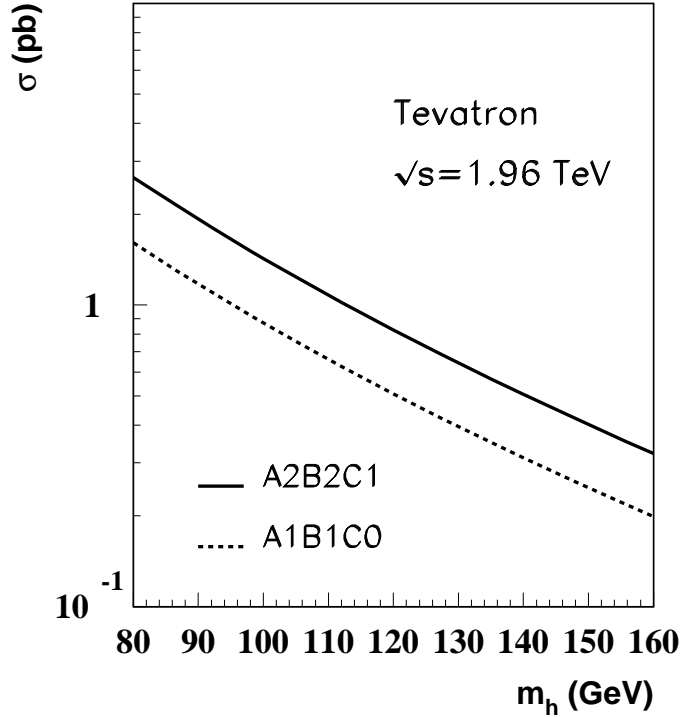}} 
\caption[]{\it The integral of the resummed cross section over all 
  $Q_T$ and rapidity for Higgs boson production at 
  $\sqrt{S} = 1.96$~TeV as a function of Higgs boson mass at 
  NLO (solid line) and LO (dashed line) accuracy.} 
\label{fig:totalcross}
\end{figure}

Integrating our $Q_T$ distributions over all $Q_T$ and rapidity $y$, 
we obtain the total cross sections as a function of Higgs boson mass 
shown in Fig.~\ref{fig:totalcross}. We may compare these cross sections 
to the values listed in Table 3 of 
Ravindran {\em et al}~\cite{Ravindran:2003um}
based on their fully inclusive calculation, without resummation.  Their 
results are obtained with different parton densities and are quoted at 
$\sqrt{S} = 2$~TeV, making the comparison somewhat imprecise.  The integral 
of our NLO resummed cross section lies between the NLO and NNLO inclusive 
results, slightly above the NLO values, for all masses shown, as would be 
expected qualitatively.  Our calculation includes the ingredients $A_g^{(1)}$ and $B_g^{(1)}$ of 
a NLO calculation with, in addition, other ingredients such as $A_g^{(2)}$ 
of the NNLO total cross section.  
\begin{figure}[ht]
\centerline{\includegraphics[width=9.0cm]{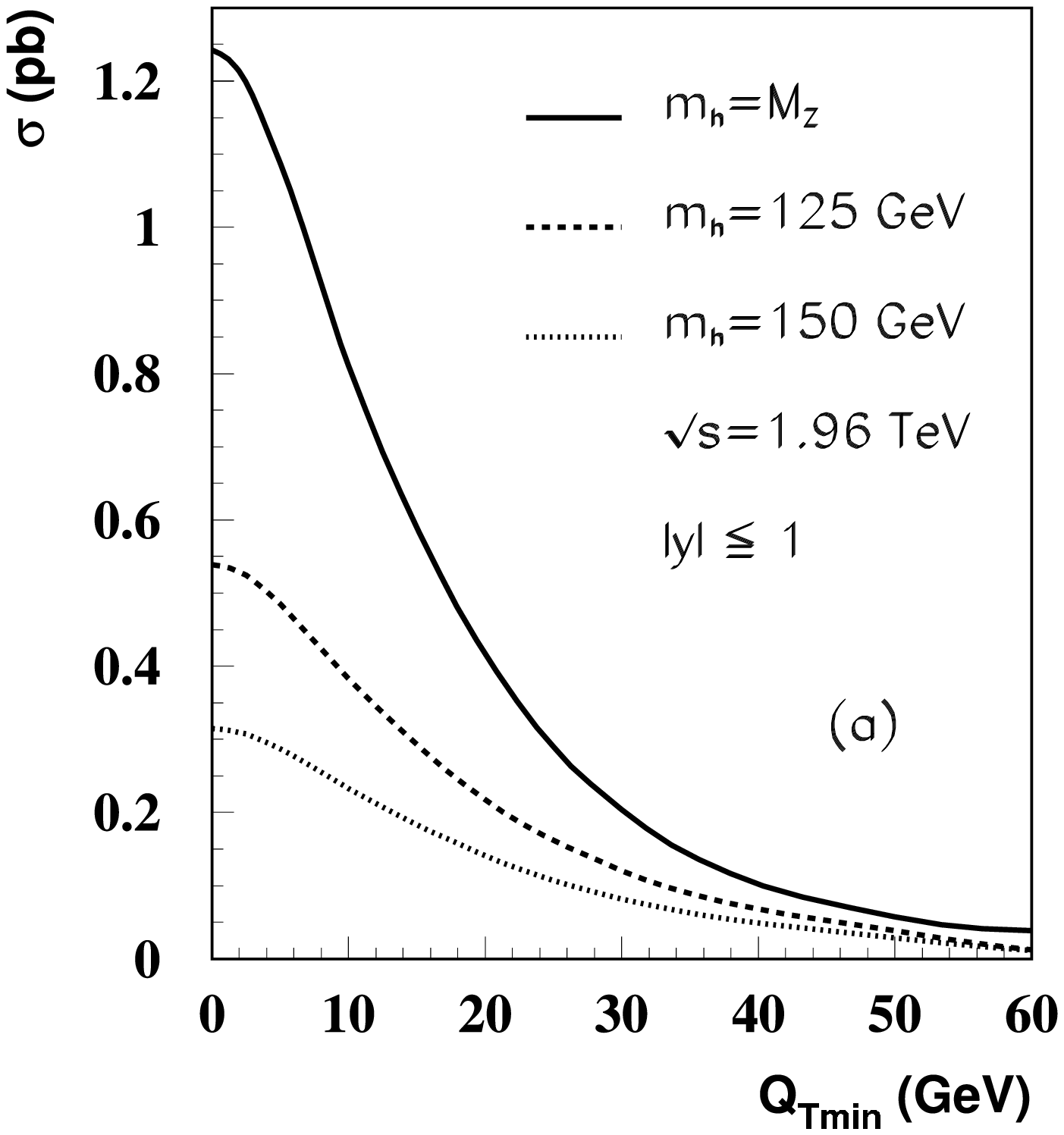}
\includegraphics[width=9.0cm]{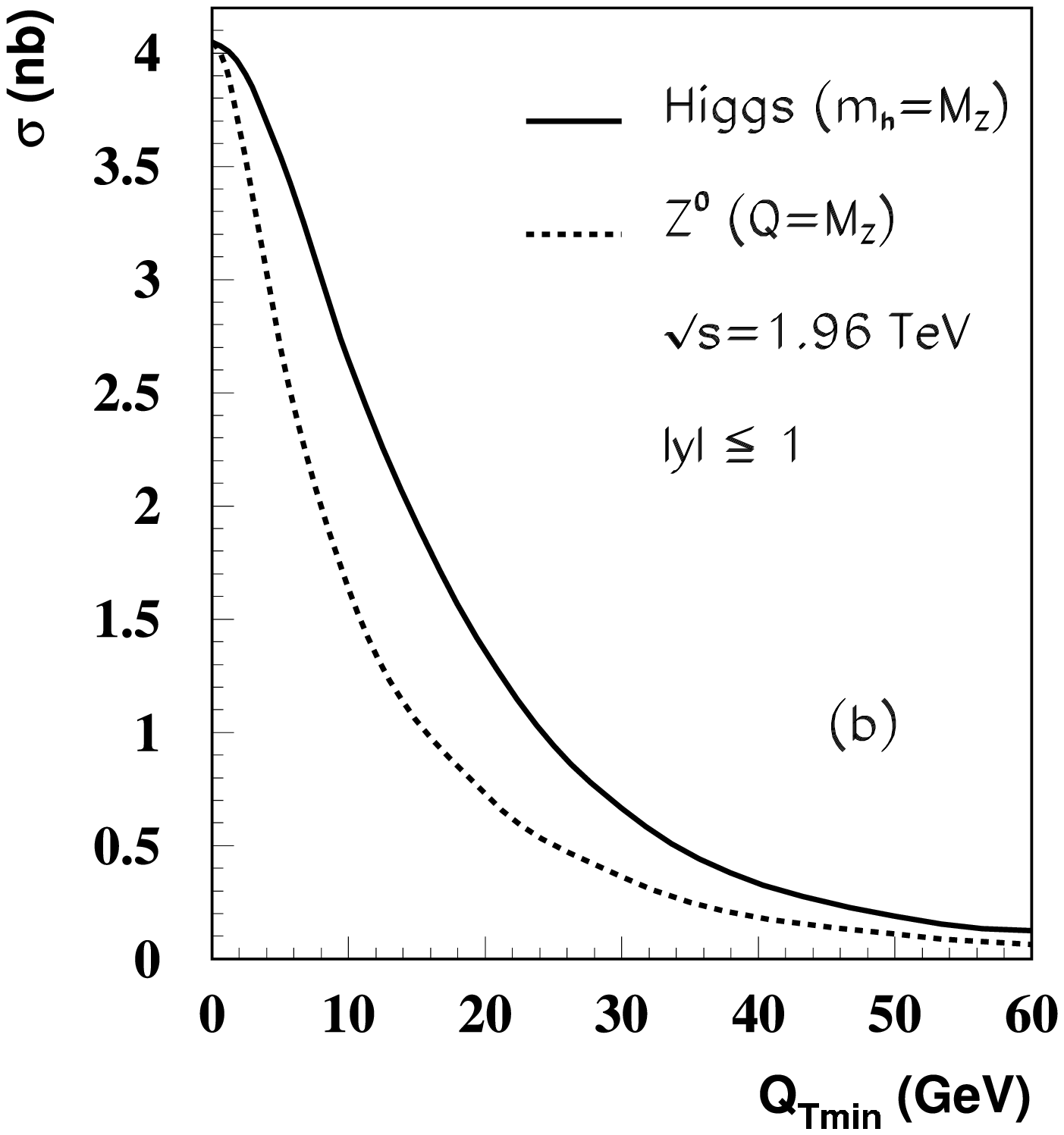}} 
\caption[]{\it (a) Cross sections for Higgs boson production at 
  $\sqrt{S} = 1.96$~TeV integrated over the ranges $Q_T > Q_{Tmin}$ 
  and $|y| < 1.0$ for three values of the mass of the Higgs boson:
  $m_h=125$~GeV, $150$ GeV, and $m_h = M_Z$.  In (b), we show analogous 
  results for $Z$ boson production and, for ease of comparison, the curve 
  for Higgs boson production at $m_h = M_Z$ rescaled so that the cross 
  section at $Q_{Tmin} = 0$ is the same as that for the $Z$ case.}
\label{fig:QTHiggsint}
\end{figure}

Comparison of the predicted $Q_T$ distributions for Higgs boson 
production at different masses of the Higgs boson  
shows that the peak of the distribution shifts to greater 
$Q_T$ as $m_h$ grows and that the distribution broadens somewhat.  
At $y= 0$, the peaks are centered at about $Q_T =$ 7.3, 8.2, 
8.7, and 9.5 GeV for $m_h = M_Z$, 125, 150, and 200 GeV, 
respectively. 
The change of the $Q_T$ distribution with $m_h$ can be 
examined quantitatively with plots of the mean value $<Q_T>$ 
and of the root-mean-square $<Q^2_T>^{\frac{1}{2}}$.  We find 
that $<Q_T>$ grows from about 19~GeV at 
$m_h = M_Z$ to about 28~GeV at $m_h = 200$ GeV.  
The curve may be approximated with a straight line over this range, 
with $<Q_T> \simeq 0.079 m_h + 12$~GeV. The root-mean-square of the 
distribution grows from about 27~GeV to about 38~GeV, reflecting 
the broadening of the $Q_T$ distribution with $m_h$.  

For comparison with Higgs boson production, we quote 
our predictions for $Z$ production: $<Q_T> = 13$~GeV and 
$<Q_T^2>^{1/2} = 20$~GeV. The difference 
$<Q^h_T> - <Q_T^Z> \simeq 6$ GeV at $m_h = M_Z$ is a manifestation 
of more significant gluonic radiation in Higgs boson production.  

In anticipation of a discussion of selections on $Q_T$ to improve 
signal to background ratios, it is instructive to examine production 
of $Z^*$ bosons, with mass greater than $M_Z$.   For $Z^*$ masses 
above $M_Z$ we assume that the dominant $q \bar{q}$ production model is 
unchanged except for the difference in mass of the $Z^*$.  We find that 
the change in the $Q_T$ distribution with mass is much less significant 
than is seen for the Higgs boson.  For example, the peak position at $y=0$ 
increases to only 3.4 GeV at $M_{Z^*} = 200$~GeV from its value of 3 GeV 
at $M_Z$.  

We introduce a new distribution for Higgs boson production that represents 
the integral of the differential cross section for all $Q_T$ greater 
than a minimum value, and for rapidity integrated over the interval 
$|y| < 1.0$. The results are shown in 
Fig.~\ref{fig:QTHiggsint}.  At $Q_{Tmin} = 10$~GeV, the integrated 
distribution has dropped to 40\% of its total for $Z$ production but 
to only 65\%, 71\%, and 74\% for Higgs boson production at $m_h = M_Z, 
125$~GeV, and $150$~GeV, respectively.  The irreducible background for 
Higgs boson decay to pairs of $W$'s arise from the QCD annihilation 
subprocess $q \bar{q} \rightarrow W W$.  This subprocess has the same 
initial state structure 
as $Z (Z^*)$ boson production.  The harder $Q_T$ spectrum for the 
glue-glue dominated Higgs boson production shows that the signal 
to background ratio can be enhanced if Higgs bosons are selected with 
large $Q_T$.  

Choices of parameters are made in obtaining our results, including the  
renormalization/factorization scale $\mu$ and the non-perturbative input.  
Our default value $\mu = c/b$ with $c = 2e^{-\gamma_E} \approx 1$ 
provides a scale that varies inversely with the impact parameter $b$.  
This selection has the virtue that logarithmic dependence on $\mu$ is 
removed from the coefficient functions 
$C$~\cite{Collins:1984kg,Berger:2002ut}.  The integration in 
$b$ space is dominated by the region near the peak in the $b$ 
distribution, $b \simeq 0.08$ for $m_h = 125$~GeV.  Using the 
conjugate relationship $Q_T \sim 1/b$, we note that the typical 
hard scale $\mu$ is therefore $< \mu> \simeq 12.5$~GeV  
($\simeq 0.1m_h$). To examine sensitivity to the selection of $\mu$, 
we consider other choices that are independent of $b$. Taking a value 
as large as $\mu = 2m_h$ produces changes in the magnitude of 
$d\sigma/dydQ_T$ at the peak position of no more than about 20\%, much 
less than the difference between the NLO and LO results in 
Fig.~\ref{fig:QTHiggs}.  
Uncertainties associated with non-perturbative physics in the region of 
large $b$ are at most 1 to 2\% depending on the size of the power 
corrections~\cite{Qiu:2000hf}. 

\section{CONCLUSIONS}

Discovery of the Higgs particle is essential to shed light on the 
mechanism of electroweak symmetry breaking. The partonic
subprocess $g+g\rightarrow h X$ dominates Higgs boson production in 
hadronic reactions when the Higgs boson mass is in the expected range 
$m_h < 200$~GeV.  The two-scale nature of the production 
dynamics, with mass $m_h$ and transverse momentum $Q_T$ both 
potentially large, and the fact that the fixed-order perturbative QCD 
contributions are singular as $Q_T \rightarrow 0$, necessitates all-orders 
resummation of large logarithmic contributions in order to obtain 
meaningful predictions for the $Q_T$ distribution  
particularly in the region of modest $Q_T$ where the cross section is 
largest.  We perform this resummation of multiple soft-gluon emission 
using an impact parameter $b$-space formalism.  The resummed $Q_T$ 
distributions are determined 
primarily by the perturbatively calculated $b$-space distributions at 
small $b$, with negligible contributions from the 
non-perturbative region of large $b$.  

In this paper, we present predictions for the $Q_T$ distributions of Higgs 
boson and $Z$ production at $\sqrt{S}=1.96$~TeV.  At the same mass, 
$m_h = M_Z$, the predicted mean value $<Q_T>$ is about 6~GeV larger for 
Higgs boson production than for $Z$ boson production.  For the Higgs boson, 
$<Q_T>$ grows from about 19~GeV at $m_h = M_Z$ to about 28~GeV at 
$m_h = 200$ GeV, and the root-mean-square $<Q_T^2>^{1/2}$ from about 27~GeV 
to about 38~GeV.   

Searches for the Higgs boson in its decay to $WW$ require an excellent 
understanding of the production characteristics of both the signal and 
backgrounds. The irreducible backgrounds arise from the QCD annihilation 
subprocess $q \bar{q} \rightarrow W W$.  Of interest to us is the expected 
$Q_T$ dependence of the ratio of signal to irreducible 
background.  In this paper we provide predictions for the transverse 
momentum distribution of the signal.  The annihilation subprocess 
$q \bar{q} \rightarrow W W$ has the same initial state structure 
as $Z (Z^*)$ boson production.  As we show, the $Q_T$ spectrum of $Z^*$ 
production is predicted to be significantly softer than that for Higgs 
boson production. We suggest therefore that a selection of events with 
large $Q_T^{WW}$ should help in improving the signal to background ratio.   
The placement of the cut requires, of course, appropriate optimization 
to maintain signal significance.  In the mass range $m_h < 135$~GeV where 
$h \ra b \bar{b}$ is the leading decay mode, the QCD subprocess 
$gg \ra b \bar{b}$ supplies the dominant background.  Since the signal 
and background are both produced from $gg$ scattering, we expect them to 
have very similar $Q_T$ dependences at $m_h = m_{b \bar{b}}$, and means 
other than selections on $Q_T$ will be necessary for background suppression 
in this mass interval.  

\section*{ACKNOWLEDGMENTS}

Research in the High Energy Physics Division at Argonne is supported 
by the United States Department of Energy, Division of High Energy 
Physics, under Contract W-31-109-ENG-38.  JWQ is supported in part by 
the United States Department of Energy under Grant No. DE-FG02-87ER40371.


\begin{thebibliography}{99}
\bibitem{NLO-LO}
We use the common but not necessarily universal terminology whereby 
resummation at next-to-leading order (NLO) accuracy means keeping terms 
through order $\alpha_s^2$ in the expansions of the functions 
$A_g$ and $B_g$ and through order $\alpha_s$ in the expansion of the 
the function $C$.  Resummation at leading-order (LO) accuracy means 
retaining terms through order $\alpha_s$ in the expansions of 
$A_g$ and $B_g$ and through order $\alpha_s^0$ in the expansion of $C$.
\end{thebibliography}
\end{document}